\documentclass[conference]{IEEEtran}
\usepackage{amsmath,amssymb,amsfonts}
\usepackage{algorithmic}
\usepackage{algorithm}
\usepackage{array}
\usepackage[caption=false,font=normalsize,labelfont=sf,textfont=sf]{subfig}
\usepackage{textcomp}
\usepackage{stfloats}
\usepackage{url}
\usepackage{verbatim}
\usepackage{graphicx}
\usepackage{cite}
\usepackage{multirow}
\usepackage{hyphenat}

\def\BibTeX{{\rm B\kern-.05em{\sc i\kern-.025em b}\kern-.08em
    T\kern-.1667em\lower.7ex\hbox{E}\kern-.125emX}}
    
\newtheorem{hypothesis}{Hypothesis}

\usepackage[dvipsnames]{xcolor}
\usepackage{listings}
\newcommand\YAMLcolonstyle{\color{red}\mdseries}
\newcommand\YAMLkeystyle{\color{black}\bfseries}
\newcommand\YAMLvaluestyle{\color{blue}\mdseries}
\newcommand{\bulletList}[1]{\begin{itemize}\item \emph{#1} \end{itemize}}

\makeatletter

\newcommand\language@yaml{yaml}

\expandafter\expandafter\expandafter\lstdefinelanguage
\expandafter{\language@yaml}
{
  keywords={true,false,null,y,n},
  keywordstyle=\color{darkgray}\bfseries,
  basicstyle=\color{black}\ttfamily\footnotesize,
  sensitive=false,
  comment=[l]{\#},
  morecomment=[s]{/*}{*/},
  commentstyle=\color{purple}\ttfamily,
  stringstyle=\YAMLvaluestyle\ttfamily,
  moredelim=[l][\color{orange}]{\&},
  moredelim=[l][\color{magenta}]{*},
  moredelim=**[il][\YAMLcolonstyle{:}\YAMLvaluestyle]{:},   % switch to value style at :
  morestring=[b]',
  morestring=[b]",
  literate =    {---}{{\ProcessThreeDashes}}3
                {>}{{\textcolor{red}\textgreater}}1     
                {|}{{\textcolor{red}\textbar}}1 
                {\ -\ }{{\mdseries\ -\ }}3,
}

\lst@AddToHook{EveryLine}{\ifx\lst@language\language@yaml\YAMLkeystyle\fi}
\makeatother

\newcommand\ProcessThreeDashes{\llap{\color{cyan}\mdseries-{-}-}}

\begin{document}

\title{MASTEST: A LLM-Based Multi-Agent System For Testing RESTful APIs}

\author{\IEEEauthorblockN{Xiaoke Han and Hong Zhu}
\IEEEauthorblockA{\textit{School of Engineering, Computing and Mathematics} \\
\textit{Oxford Brookes University}\\
Oxford, UK \\
EMail: hzhu@brookes.ac.uk}
}

\maketitle

\begin{abstract}
Testing RESTful API is increasingly complicated but indispensable to quality assurance of cloud-native applications. This paper reports a multi-agent system called \emph{MASTEST} that combines LLM-based intelligent agents and programmed agents to automate REST API testing. They form a complete tool chain covering the whole workflow of REST API test with API specification in the OpenAPI Swagger format as the input. It also incorporates human testers in the process to review and correct LLM generated test artefacts to control the quality of testing activities. MASTEST is evaluated on two LLMs, GPT-4o and DeepSeek V3.1 Reasoner with five public APIs. Its performances on various testing activities are measured by a wide range of metrics, including adequacy and coverage metrics, the syntax and data type correctness of generated test scripts, the usability of LLM generated test cases and scripts, as well as the bug detection ability. Experiment results demonstrated that both DeepSeek and GPT-4o achieved a high overall performance but had strengths and weaknesses on different testing activities. MASTEST generated test cases achieved 94\% and 98\% unit test coverage and 79\% and 78\% system test coverage for GPT-4o and DeepSeek respectively in comparison with human designed test cases. The generated test scripts maintained 100\% syntax correctness and only required minimal manual edits for semantic correctness. The generated test scripts contain assertions on the expected status code as well as contents in the response messages. They are highly capable of detecting bugs in the REST APIs. Experiment data shows that the bug detection rates are between 2.13 to 4.50 per operation. These findings indicate that MASTEST is highly efficient and effective. 
\end{abstract}


\section{Introduction}\label{sec:Introduction}

REST APIs have been widely adopted for cloud-native computer applications as the program interface for RESTful web services as well as internally between components in the microservices architecture. However, the complexity of the structure and dynamic behaviour of cloud-native applications imposes a grave challenge to their testing and quality assurance. In the current best practice, two approaches to testing REST APIs have been widely adopted in the IT industry. 

The first is GUI tool supported manual testing. In this approach, testers manually prepare test data, submit the test data to the service under test and receive the responses from the services via interactive operations of a testing tool's GUI. The most commonly used such testing tools include \emph{Postman}\footnote{Postman: \url{https://www.postman.com/}} and \emph{Swagger UI}\footnote{Swagger UI: \url{https://swagger.io/} \label{fnt:Swagger}}. 

The second approach is test script/code based automated testing. In this approach, tests are automated via running manually written test code or scripts based on test frameworks developed for various programming languages, such as \emph{Pytest}\footnote{Pytest: \url{https://docs.pytest.org/en/stable/}\label{fnt:PyTest}} for Python and \emph{JUnit}\footnote{JUnit: \url{https://junit.org/}} and \emph{TestNG}\footnote{TestNG: \url{https://testng.org/}} for Java. These test frameworks provide general-purpose testing capabilities that can be applied to API testing as well as unit and integration testing. They offer features like organising test cases and test data, setting up and tearing down test environments, checking the correctness of service responses, and reporting test results, etc. They also support the collaborations among large teams. Being programming language based and thus highly flexible and expressive, these frameworks allow integration with external data sources, databases, or other testing utilities to achieve a high degree of test automation. These make them suitable for complex test workflows and large-scale automated testing. They are most widely used in industry. 

However, both approaches require manual design of test cases, preparation of test data. In the framework approach, although test code/script automates the execution of testing process, it requires programming skills to write test code/scripts, which demands a significant amount of effort. It is highly desirable to further automate the testing of REST APIs. 

In the past decade, a large amount of research has been reported in the literature to automate the testing of REST APIs; see \cite{golmohammadi2023testing} for a recent survey. Among the most well-known are EvoMaster \cite{arcuri2017restful, arcuri2019restful}, RESTler \cite{atlidakis2019restler}, QuickREST \cite{karlsson2020quickrest}, RestTestGen \cite{viglianisi2020resttestgen}, RESTest  \cite{martin2021restest}, RestCT \cite{wu2022combinatorial}, ARTE \cite{alonso2022arte}, Morest \cite{liu2022morest}, foREST \cite{lin2023forest}, FuzzTheREST \cite{dias2024fuzztherestConf}, DeepREST \cite{corradini2024deeprest}, RAFT \cite{saha2025rest}, etc. 

These methods are mostly based on the specification of the REST API under test, typically in the OpenAPI specification format, also known as \emph{Swagger}. From such a specification of a REST API under test, algorithms have been advanced to extract the list of resources handled by the service, the formats of service requests and responses, and the list of operations on these resources that the service offers and the inter-dependences between this operations. Such information are then used to define the input space of test cases, generate test sequences and check service responses algorithmically, and sometimes heuristically. 

However, despite the intensive research efforts, automated REST API test is still not a common practice in the IT industry. ``\emph{There is still a lot of research work that is needed to be carried out in this domain}'' \cite{ golmohammadi2023testing}. 
We believe that the main drawback of the algorithmic approach to REST API test automation is that the semantic information contained in the natural language descriptions of API operations and the words used to name data types, operation parameters and schemas of response messages cannot be effectively captured and utilised in the generation of test cases and data and automation of test oracle. With the rapid advances of large language models (LLMs), machine learning (ML) techniques in natural language processing (NLP) offers a promising new solution to this problem.

In fact, recently researchers have explored LLMs' capability in understanding natural language descriptions to assist in API testing. For instance, empirical studies have demonstrated the applicability of LLMs to generate diverse input parameters \cite{alonso2022arte, le2024kat, kim2024leveraging}, to interpret the upstream and downstream dependencies between API operations and restrictions on API responses \cite{corradini2024deeprest, huynh2024using}, and to generate test code from program source code \cite{rincon2025llm}, business requirements \cite{pereira2024apitestgenie} and API specifications \cite{sri2024automating}. However, they have failed to achieve their goal of complete automation of REST API test due to the the hallucination and imperfect performance problems of LLMs. There is still a wide gap to reach a practical solution. The following key research problem still remains open: 

\emph{Given the fact that LLMs inevitably suffer from imperfect performance, how can LLMs be used to build a REST API test automation tool that achieves required quality of testing for practical uses}?  

In this paper, we propose a role-based multi-agent system, called \emph{MASTEST}, which integrates a group of agents of different kinds to play different roles. These agents are either ML-based and empowered by a LLM, or algorithmic and implemented in a programming language, or human beings supported by GUI interfaces to enable human testers in the loop. Each agent is designated to a specific role to perform a specific testing activity. Collectively, they cover the whole workflow of testing REST APIs and collaborate through sharing data and states of testing tasks. 

The design goal of MASTEST is to reduce the workload on human testers with controllable quality of testing tasks instead of full automation or complete autonomy. To achieve this goal, LLM-based intelligent agents are employed to conduct labour intensive but creative tasks that cannot be well performed algorithmically with satisfactory accuracy and precision, like the generation of test scenarios and the generation of test scripts that contains both service requests and assertions on response correctness. Programmed agents are assigned to routine tasks that can be performed well algorithmically with high accuracy and precision or invoking existing software tools, such as parsing REST API specifications in the Swagger format, running test code to submit test request sequences to web services under test, receiving response messages from web services, checking the correctness assertions and calculating test coverage. Human testers participate in the workflow to perform quality control tasks such as reviewing the output generated by LLM-based agents and correcting their errors. These quality control tasks are also supported by metrics automatically produced by other types of agents. 

The paper is organised as follows. Section \ref{sec:DesignAndImp} presents the design and implementation of the MASTEST system. Section \ref{sec:Evaluation} reports the experiments that evaluate the system. Section \ref{sec:RelatedWorks} discusses related work and compares them with MASTEST. Section \ref{sec:Conclusion} concludes the paper with a summary of the main contributions of the paper and a discussion of the future work. 

\section{Design And Implementation of The System}\label{sec:DesignAndImp}

This section presents the design and implementation of the MASTEST system. We will outline the overall workflow and the architecture of the multi-agent system, present the designs of intelligent agents that employ LLMs to realise their functionality, and the design of GUIs that supports human involvement in the testing process. Finally, we briefly describe the implementation of the system. 

\subsection{Workflow and System Architecture}\label{sec:WorkflowAndArchitecture}

The architectural design of MASTEST system is based on a careful analysis of the workflow of the current best industrial practice to answer the following key questions in the design of a role-based multi-agent system.  
\begin{enumerate}
\item What are the agent roles that should be included in the system? 
\item How should the functionality of each agent role be implemented? And,
\item How will agents of different roles collaborate with each other? 
\end{enumerate}

To answer the first two questions, a set of testing tasks currently performed by human testers in the framework based approach are identified. These tasks cover the whole workflow of REST API test process. Each task is then assigned to one agent role in the multi-agent system. To enable collaborations between these agents, the analysis also identified the input and output of each task as they occur in the current best practice as well as how one task's output is passed to the other tasks as input. 

To decide how an agent role should be implemented, we applied the following design rules to three types of agents as stated in Section \ref{sec:Introduction}. 

\begin{itemize}
\item \emph{Programmed agents}, which are algorithmically implemented by writing traditional program code or invoking existing software engineering tools. They are utilised as much as possible for testing activities and tasks that are feasible to do so. 
\item \emph{Intelligent agents}, which are implemented by querying a LLM with appropriate prompts. They are employed for testing activities that are constructive, generative, creative, and labour intensive, but infeasible or too expensive to be implemented algorithmically, under the condition that the activity is perceived to be feasible to be solved by LLMs within a tolerable error rate. 
\item \emph{Human agents}, who are supported by graphical user interfaces to perform a task and interact with other agents. Each GUI enables the tester to perform one testing task manually, if the task is infeasible for both algorithmic and intelligent agents and crucial to quality control. 
\end{itemize}

It is worth noting that assessing the feasibility of a role for implementation by using a LLM is a challenge, because it is hard to predict accurately the performance of a LLM on a task if performance evaluation experiments on the specific task has not been conducted or reported in the literature. In the design of MASTEST, we assess informally the feasibility of each role assigned to an intelligent agent based on the following hypothesis. 

\begin{hypothesis}(\emph{LLM Capability})

\textit{When the workflow of a LLM-based multi-agent system follows the same process widely used in the current practice and the same types of software artefacts are used and generated, a large volume of data from software engineering practice is available to train LLMs. Consequently, LLMs could be highly capable of performing the same generative tasks in the workflow, i.e. highly capable of understanding the input artefacts and generating the corresponding output artefacts.} 
\end{hypothesis}

In our case, the \emph{LLM Capability} hypothesis means that it is reasonable to assume that LLMs are highly capable of understanding specifications of APIs in Swagger format, highly capable of understanding and generating natural language descriptions of test case in the form of operation scenarios, and highly capable of understanding and generating test scripts in PyTest library of Python from scenario descriptions of test cases. Consequently, tasks that rely on these capabilities should be feasible for LLMs to perform, thus they are suitable for intelligent agents. 

When assigning a task to a human agent in the workflow, a question is that: \emph{whether the human agent can perform the task effectively and efficiently so that his/her workload can be reduced and their productivity can be improved}? We assume that the following hypothesis holds. 

\begin{hypothesis}(\emph{Human Adaptability})

\textit{When the workflow follows the same process widely used in the current practice and the same types of artefacts are used and generated, it is easy for human users to adapt to the new environment of LLM-based multi-agent system  with little training. Their past experiences, knowledge and skill of conducting constructive and creative tasks can be easily applied to the new role of reviewing LLM generated artefacts and correcting the errors made by LLMs.}
\end{hypothesis}

It is reasonable to assume that reviewing a software artefact and correcting errors in such an artefact can be performed more efficiently than manually producing such an artefact, especially when the number of errors produced by LLMs are small. Therefore, the \emph{Human Adaptability} hypothesis implies that one can predict the improvement in the productivity of software engineers if their role is shifted from producing a type of software artefact to reviewing and correcting the same artefact generated by a LLM. In other words, the goal of our architectural design is achievable by following the role assignment rules stated above. 

\begin{figure}[htb]
\centering
\includegraphics[width=2.5in]{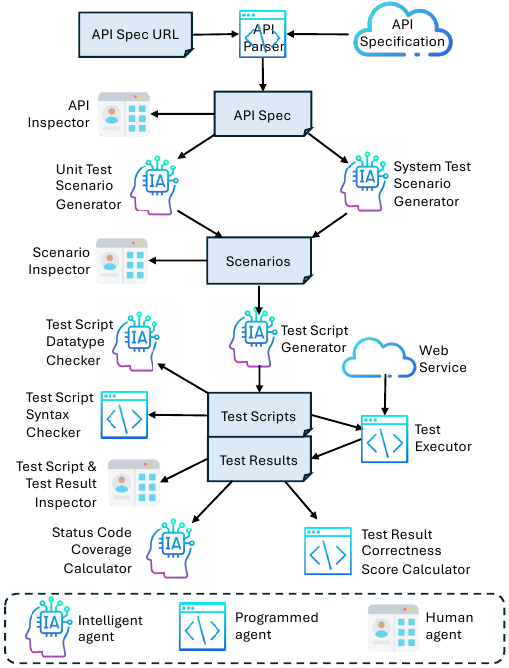}
\caption{Workflow and MASTEST Architecture.}
\label{fig_1}
\end{figure}

Fig. \ref{fig_1} presents the system architecture and illustrates the roles assigned to different types of agents and how they are integrated through sharing data to support the workflow and interact with human testers.

As shown in Fig. \ref{fig_1}, MASTEST system consists of the following agents. 
\begin{itemize}
\item \emph{API Parser}. It is a programmed agent. It loads an API specification in Swagger$^{\ref{fnt:Swagger}}$ format and parses it into a list of API methods and various other parts to enable the analysis of the API specification in a smaller granularity in the downstream of the workflow, such as to generate test scenarios targeting an individual method, or a subset of methods. 
\item \emph{Unit Test Scenario Generator}. Given the specification of an API method, it invokes a LLM to generate a set of test scenarios for testing the method. Each test scenario describes a usage of the operation to be tested in natural language so that it can be reviewed by a human tester and used as the input to generate test scripts. 
\item \emph{System Test Scenario Generator}. Given a set of API methods as the input, it invokes a LLM to generate a set of system test scenarios described in natural language. Each system test scenario typically involves a sequence of API method calls. The human tester can select a subset of API methods to generate system test scenarios that only call the selected API methods, or let the LLM-based agent to generate from all methods. The generated system test scenarios include both meaningful usages (i.e. a valid sequence of API method invocations) and faulty usages (i.e. an invalid sequence of API method invocation). Both of them are used to generate test scripts to perform system level tests. 
\item \emph{Test Script Generator}. Given a test scenario description and the specifications of the API methods involved in the scenario as the input, it invokes a LLM to generate test scripts in PyTest framework. Note that testing a service on a scenario normally requires a number of test cases to cover various behaviours of the service in the scenario. Each test case is normally represented as a test method in PyTest framework that include code to request the service and code to check the correctness of the responses from the service. Therefore, a LLM-generated test script for one scenario typically contains multiple test methods that each represents a test case.
\item \emph{Test Script Data Type Checker}. Given a test script and the specifications of the API, it invokes a LLM to check if the datatype in the test script in Python matches the data types of the invoked API methods. Being a dynamic typing programming language, verifying data type correctness is a non-trivial task. There is no existing tools available for our purpose. 
\item \emph{Test Script Syntax Checker}. This is a programmed agent that calls the Python library to check if a test script is syntactically valid. When an error is detected, the line(s) of test scripts is marked so that the human tester can make corrections by editing the test script. 
\item \emph{Test Script Executor}. It is also a coded agent. Given a test script, it executes the test script, which typically include a sequence of statements that send service request(s) with specific test data as the parameters and save the received response message(s) from the service into variable(s), and a sequence of assertion statements that check if the response message is correct in status code and also in contents. 
\item \emph{Status Code Coverage Calculator}. It is a LLM-based agent but also include program code to fulfil its functionality. Note that each REST API request returns a status code to indicate whether the service was executed correctly, and the error code if not. For example, status code 200 typically means that the invocation of the API method was terminated normally. LLM-generated test scripts for one scenario typically contains multiple test methods. Each method represents a test case and returns a status code when it is executed. Status Code Coverage is a metric that measures how thoroughly the set of test cases test the service on a scenario by the proportion of all possible status codes of the scenario covered by the status codes returned by the set of test cases. It is a test adequacy criterion widely used in industry for human written test scripts and also used in the research on REST API testing methods; see, for example, \cite{martin2019test}. This agent employs LLM to analyse the API specification as well as the test reports returned by the executions of test methods for the test scenario to work out the set of all possible status codes and the set of returned status codes. The agent then applies a coded algorithm to calculate the coverage score.
\item \emph{Test Result Correctness Score Calculator}. It is also a coded agent. It collects the test results generated by test scripts after the Test Script Executor executes them. The correctness of the responses from the service under test is determined by the assertion statements in test scripts. When the status code in the response from a service request mismatches the expected, an error is detected. Moreover, a test script typically also contains a number of assertions on the contents of the response message. When an assertion fails, it also detects an error. Given a test scenario and a set of test methods (i.e. test cases) for the scenario, this agent collects the correctness data from the test results for each test method and calculate the correctness score as the proportion of test cases that are correct, which is formally defined by the following formula. 

\begin{equation}
Cor_{Sem}(scn) =\frac{|\{t \in T(scn)|Correct(t)\}|}{|T(scn)|}\label{equ:CorrectnessScore1}
\end{equation}
where $T(scn)$ is the set of test cases in the form of test methods for testing in scenario $scn$. $Correct(t)$ is the predicate that all assertion statements in test method $t$ are true when it is executed. $|X|$ is the number of elements in the set $X$ , e.g. $|T(scn)|$  is the number of test cases in $T(scn)$. 

It also calculates the overall correctness score for the whole API, which is the average correctness score over all test scenarios and formally defined by the following formula. 
\begin{equation}
Cor_{Sem}(api) =\frac{1} {|T(api)|}\cdot {\sum_{ts \in T(api)}Cor_{sem}(ts)} \label{equ:CorrectnessScore2}
\end{equation}
where $T(api)$ is the set of test scripts generated by LLM for the API $api$. $|T(api)|$  is the number of test scripts in $T(api)$. $Cor_{sem}(ts)$ is the correctness score for test scenario $ts$ represented in the form of a test script $ts$ in $T(aps0$. It is calculated according to Equ. (\ref{equ:CorrectnessScore1}). 
\end{itemize}

To include humans in the loop, graphic user interfaces (GUIs) are designed to enable human testers to review the output generated by LLMs, make corrections if the output is incorrect, add additional scenarios and test scripts that LLM missed, and control the process by selecting the specific tasks to be performed by the intelligent agents. As shown in Fig. \ref{fig_1}, MASTEST offers three types of GUI windows (implemented as a web page) that are associated to the key artefacts of the test process, i.e. API operations, test scenarios, and test scripts together with their execution results. They are called \emph{API Inspector}, \emph{Scenario Inspector} and \emph{Test Script and Test Result Inspector}, respectively. 

The workflow is designed to break down the entire testing process into multiple stages, allowing human testers to review outputs at each stage and ensure the quality of the outcomes of each stage before moving to the next. This approach mitigates the problem of error accumulation and amplification, which inevitably occur if software engineering artefacts are generated by LLMs autonomously without quality control. 

The workflow is also designed to break down each stage into multiple smaller activities. For example, generating the unit test scenarios for the whole system is decomposed into generating unit test scenarios for each individual API operation. The generation of all test scripts for the whole system is decomposed into the generation of test scripts for each scenario. It improves the overall reliability and scalability of the automated testing process by avoiding very long prompts to LLMs. 

In comparison with the current practice of API test, all labour-intensive, time-consuming generative activities are now delegated to LLM-based intelligent agents and routine and tedious tasks assigned to coded agents. The focus of human testers is shifted to reviews and corrections of LLM generated artefacts and assisted by other agents. It is worth noting that, in the current best practice, those test artefacts generated by LLM-based agents in the MASTEST system are not only manually written, but also checked, reviewed and corrected manually either by the engineers themselves or by more senior testers, even by running review workshops. Therefore, such manual review and correction tasks are not additional works in comparison with the current practice of REST API test. Thus, MASTEST can be significantly reduce the workload on human testers and improve their productivity while maintaining the quality of testing tasks. 

\subsection{Intelligent Agents}

The intelligent agents in the system rely on prompts to invoke LLMs to realise their functions. Although the prompt for each type of agent has its specific content depending on the function and input/output data, all prompts follow the same structure of dual-roles, which include:
\begin{itemize}
\item \emph{A system prompt} provides the background information and defines the role that the agent is playing. 
\item \emph{A user prompt} provides the required input data as the context and clear instructions on the requirements of the output. 
\end{itemize}  

\begin{figure}[htb]
\centering
\includegraphics[width=3.4in]{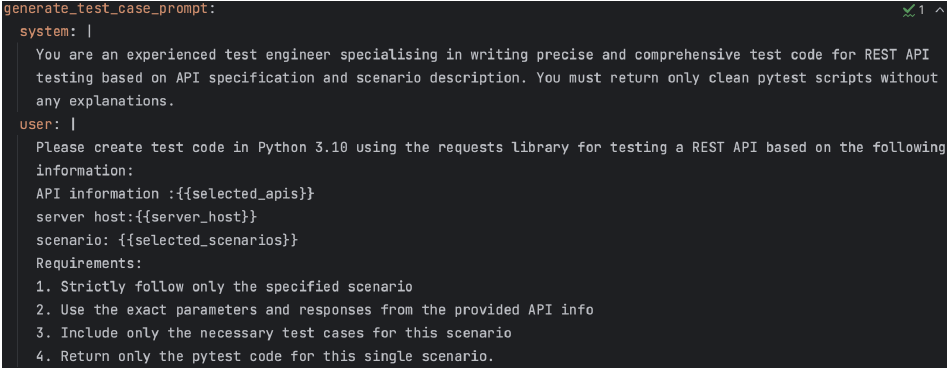}
\caption{Prompt Template for Generating Test Scripts.}
\label{fig_2}
\end{figure}

Fig. \ref{fig_2} shows the prompt template for generating test scripts. The prompt templates for other agents can be found in the technical report \cite{han2025mastest}. %Appendix \ref{sec:PromptTemplate}.

During execution, variables in each prompt template are replaced with actual data by the agent. The values of these variables form the context of the prompts. 

The key question that we have confronted in the design of prompts is: \emph{what should be included in the prompt as the context?} We have followed the principle that \emph{the data and information that human testers need to produce the output or to review the output should also be included in the prompt as the context}. In general, we assume that the following hypothesis holds for other application domains as well. 

\begin{hypothesis}(\emph{Human and LLM Homogeneity})

\textit{The data and information that a LLM requires to perform a generative task is the same data and information that human users need to produce the output, which is also the same for the human users to validate and verify the output.}
\end{hypothesis} 

Table \ref{tab_1} gives the variables in the prompt templates used by various LLM-based agents, where API method details include \emph{URI path}, \emph{method name}, \emph{summary}, \emph{input parameters} and \emph{response}. 

\begin{table}[ht]
\caption{Variables in Prompt Templates}
\label{tab_1}
\centering
\begin{scriptsize}
\begin{tabular}{|p{30mm}|p{49mm}|}
\hline
\textbf{Agent} &\textbf{Variables}\\\hline
Unit test scenario generator &API method details\\\hline
System test scenario generator &A list of API method details\\\hline
Test script generator &API method details, Scenario description, Service host URL\\\hline
Test script data type checker &Scenario description, API method details, Test script\\\hline
Status code coverage calculator &Scenario description, API method details, Test script, Execution results
\\\hline
\end{tabular}
\end{scriptsize}
\end{table}

Note that the \emph{Status Code Coverage Calculator} uses two different prompt templates depending on whether the execution of the test script is successful. 

\subsection{Interface Design}\label{sec:InterfaceDesign}

\begin{figure}[htb]
\centering
\includegraphics[width=3.4in]{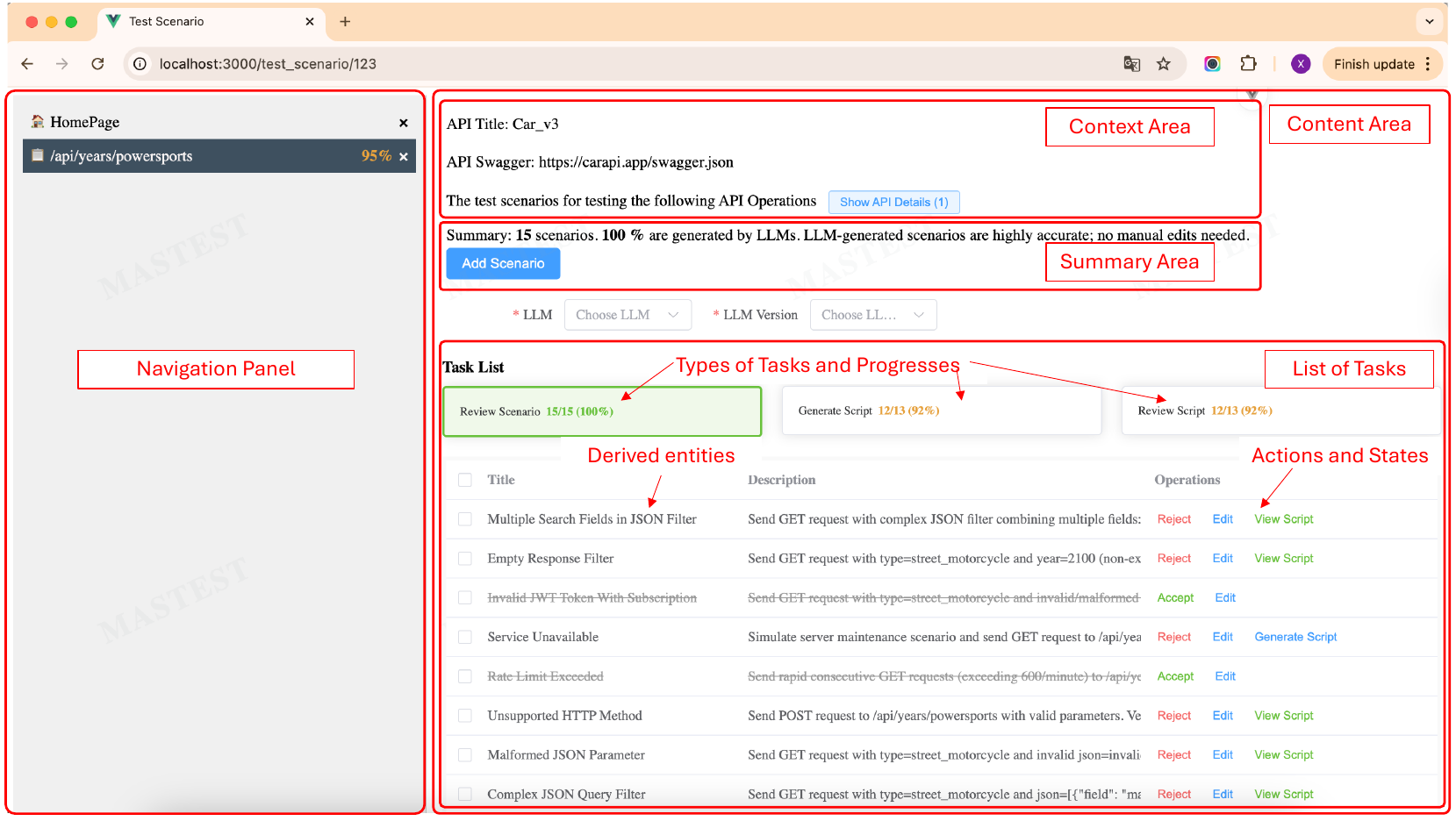}
\caption{Interface of API Operation Inspector.}
\label{fig_3}
\end{figure}

The graphic user interface plays a crucial role in supporting human testers to control the testing process in the workflow. The windows of the GUI are divided into two areas; on the left-hand-side is the \emph{navigation panel} and, on the right is the \emph{content area}. An example is shown in Fig. \ref{fig_3}, which is the graphic user interface of the Scenario Inspector for reviewing, editing and approving unit test scenarios generated for an API operation. 

The \emph{navigation panel} lists the entities of the project generated so far and links to a web page that displays its details, state and actions to be performed. Each item in the navigation panel is associated with an icon to indicate its type, the name of the entity, and a percentage number to indicate the progress in its completion of construction. The details of each entity is represented by a different tab/page of the GUI window. A click at an item in the navigation panel will display the corresponding tab/page in the web browser window. With the development of the testing project, a new entity could be added to the navigation panel once it is created either by an intelligent agent or manually by the human user. An existing one could also be removed from the navigation panel if the entity is removed by the human tester. It provides an overview of the progress of the testing project, an overview of the set of test artefacts of the project, and also enables navigations among them easily. Table \ref{tab:GUITopLevel} lists the types of items in the navigation panel and the contents of corresponding web pages that each links to. 

\begin{table}[ht]
\caption{Links Between Navigation Panel and Web Pages}
\label{tab:GUITopLevel}
\begin{scriptsize}
\begin{tabular}{|l|p{5.7cm}|}
\hline
\textbf{Nav Panel Item} &\textbf{Contents of Linked Web Page}\\\hline
Project homepage &It gives the name and URL of the API specification\\\hline 
API specification &It lists the operations of the API and shows their states in the project\\\hline 
Unit test scenarios &Each lists the set of unit test scenarios generated for one API operation and shows their states\\\hline  
System test scenarios &Each lists the set of scenarios generated for all operations of the system or a selected subset of operations and shows their states\\\hline 
Operational scenarios&Each lists the subset of system test scenarios that involves a specific API operation and shows their states\\\hline 
Test scripts &Each lists the set of test scripts generated for one test scenario, shows their states and populated with test results after the test script is executed. \\\hline
\end{tabular}
\end{scriptsize}
\end{table}
 
The \emph{content area} displays the content of the corresponding entity, which is referred to as the \emph{subject} of the web page/tab. On the top of the area is the \emph{context area}, which provides the context information of the subject. For example, Fig. \ref{fig_3} shows the web page for the unit test scenarios of an API operation. The API operation is the subject of this page. The context area includes the \emph{API title}, the \emph{API URI}, the \emph{API operation under test}, which is hidden in the figure but can be shown by a click on the \emph{Show API Details} button. The context area provides the data to the variables in the prompt template used by the intelligent agents to perform testing activities on the entity. It also enables the human user to perform review and validation activities on the entity without switching to other web pages/tabs for context information. 

Below the context area is the \emph{summary area}, which provides the state of the subject entity with three types of metrics on (a) the size of the subject entity, (b) the progress of test activities on the subject entity, and (c) the quality of the subject entity. The metrics for various types of the subject entities are given in Table \ref{tab_2}. 

For example, Fig. \ref{fig_3} shows the total number of unit test scenarios for the API operation, and the numbers of them generated by the agents, added by the user manually, and modified by the user, respectively. 

\begin{table*}[htb]
\caption{Derived Entities and Metrics Associated to Each Subject Entity}
\label{tab_2}
\centering
\begin{scriptsize}
\begin{tabular}{|p{10mm}|p{22mm}|p{24mm}|p{33mm}|p{65mm}|}
\hline
\textbf{Entity} &\textbf{Derived Entities}&\textbf{Size Metrics} &\textbf{Progress Metrics} &\textbf{Quality Metrics}\\\hline
\multirow{2}{10mm}{API Spec}&API Ops &\# API Ops &\# API Ops unit test completed, ~~~~~\# API Ops system test completed& \\\cline{2-5}
&System test scenarios & \# System test scenarios &\% Scenarios reviewed &\% Scenarios Accepted, \#  scenarios edited \\\hline
\multirow{2}{10mm}{API Op} &Unit test scenarios&\# Unit Test Scenarios &\% Scenarios reviewed & \% Scenarios accepted, \# Scenarios manually added, \# Scenarios edited \\\cline{2-5}
&System test scenarios &\# System test scenarios &\% Scenarios reviewed &\% Scenarios accepted, \# Scenarios manually added, \# Scenarios edited \\\hline
Test Scenario &Test scripts (cases) &\# Test scripts &\% Test scripts reviewed, \% Test scripts executed &\% Test scripts accepted, \# Test scripts manually added, \# Test scripts edited, \# Test scripts failed, \# Syntax errors, \# Data type errors, \# Semantic errors\\\hline
%Test Script&Lines of Code &\# Lines of code &isExecuted &\# Syntax errors,\# Data type errors, \# Semantic errors
%\\\hline
\end{tabular}
\end{scriptsize}
\end{table*}

The lower part of the content area is the \emph{task list area}. This area lists the entities derived and generated from the subject entity and the corresponding testing activities to be performed on them. For example, Fig. \ref{fig_3} shows a list of scenarios generated by the Unit Test Scenario Generator agent for the subject entity, i.e. an API operation. These test scenarios are the unit test scenario that the API operation should be tested on. Above the list of derived entities shows the types of manual testing activities to be performed on each item in the list and the progress so far. For example, for each scenario in the list, the testing activities to be performed are: (a) to review, validate and make corrections (if needed) of the scenarios, and then (b) to invoke the intelligent agent to generate test scripts from the scenario descriptions, and finally (c) to review the test scripts and execute them. In the list of derived entities, buttons are associated with each item for the user to perform manual testing activities and reflect the item's state. For example, as shown in Fig. \ref{fig_3}, the following three buttons are associated with each scenario in the list. 
\begin{enumerate}
\item \emph{Reject/Accept}. It is initially in the form of \texttt{Reject}. A click on \texttt{Reject} will remove the scenario from the list and the scenario name and description will be displayed in the strike-through format, and then button will change to the form \texttt{Accept} to enable the removed scenario to be revoked back to the list of scenarios. 
\item \emph{Edit}. It is to manually modify the scenario. A click on this button, a pop-up window will show to enable the manual editing of the scenario. 
\item \emph{Generate/View Script}. It is initially in the form of \texttt{Generate Script}. A click of the button means that the review of the scenario is completed and it will invoke the intelligent agent to generate test scripts for the scenario. Then, the button will be changed to the form \texttt{View Script}. A click on \texttt{View Script} will open and jump to the window to display the test scripts derived from the scenario for reviewing and executing the test scripts. 
\end{enumerate}

The human tester should work on the list of derived entities from top to the button one by one until the progresses are 100\%. Moreover, the human tester should also review the completeness of the list of derived entities. If incomplete, additional entities, for example, additional scenarios, can be added manually using the \emph{Add Scenario} button. Each action taken by the human tester is reflected in the summary area. Table \ref{tab_2} gives the type of derived entities for each subject entity. Screen snapshots of the windows for other types of entities can be found in the technical report \cite{han2025mastest}. %Appendix \ref{sec:Screenshots}.

This design of GUI followed the \emph{task list metaphor}, which forces the human tester to review, validate, and correct errors in fine granularity progressively and systematically. It allows the human testers to easily track progress, navigate between steps, and interact with intelligent agents without manually entering complicated prompts. By showing tasks visually and providing clear progress indicators, the interface improves usability and helps maintain human control in the LLM-assisted automation process. It also supports regression testing when the API is revised by reusing existing scenarios and test scripts that are not affected by the API changes and adding or modifying the scenarios and test scripts affected by the API changes. It is worth noting that the specific data and information displayed on screen for human tester to perform a specific type of test activity follows the hypothesis of \emph{Human and LLM Homogeneity}. This minimise the need to navigate among different web pages to find data and information. 

\subsection{Implementation}

\begin{figure}[htb]
\centering
\includegraphics[width=2.5in]{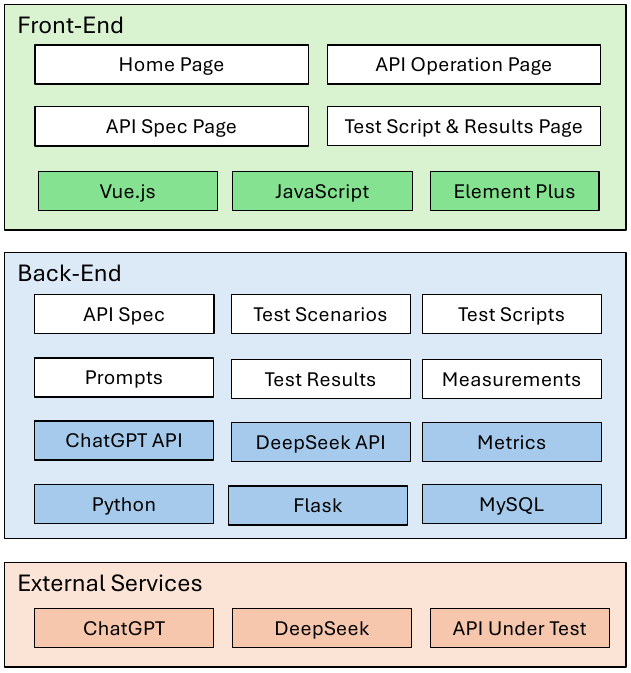}
\caption{System Structure}
\label{fig_4}
\end{figure}

The system has been implemented with a front-end/back-end separation structure as shown in Fig. \ref{fig_4}. 

The front-end is built with \emph{Vue 3}
\footnote{Vue: \url{https://vuejs.org/guide/introduction.html}}, 
\emph{JavaScript}
\footnote{JavaScript: \url{https://developer.mozilla.org/en-US/docs/Web/JavaScript}}, 
and \emph{Element Plus}
\footnote{Element: \url{https://element-plus.org/en-US/component/overview.html}} component libraries. It implements the GUI described in subsection \ref{sec:InterfaceDesign}.

The back-end is implemented in \emph{Python 3}\footnote{Python: \url{https://www.python.org/doc/}} with \emph{Flask}\footnote{Flask: \url{https://flask.palletsprojects.com/en/stable/}}. It manages the interactions with LLMs, processes API specifications, and performs test scenario and script generation, and collects data of human interactions and calculates metrics. A database is used to store API operation details, scenarios, scripts, execution results, and metric values. The database is managed, and queries processed by using \emph{MySQL}\footnote{MySQL: \url{https://dev.mysql.com/doc/}} and deployed with \emph{Docker}\footnote{Docker: https://docs.docker.com/}.
 
\section{Evaluation}\label{sec:Evaluation}

In this section, we will report the evaluation experiments with the system. 

\subsection{Experiment Setup}

To evaluate the feasibility of using LLMs for automated RESTful API testing, five public APIs were selected at random as a small benchmark according to the following criteria. (a) The API is specified in Swagger format, (b) Real cloud native application, and (c) The web service is deployed and accessible. 
Table \ref{tab_3} lists the selected APIs. The Methods column lists the HTTP2 methods used by the API, the \#Ops column gives the number of API operations included in the API, and the URL column gives the URL address of the web service. 

\begin{table}[htb]
\caption{The Benchmark Dataset}
\label{tab_3}
\centering
\begin{scriptsize}
\begin{tabular}{|l|l|c|l|}
\hline
\textbf{Name} &\textbf{Methods} &\textbf{\#Ops} &\textbf{URL} \\\hline
Car &Get, Post &24 &\url{https://carapi.app}\\\hline
Petstore &Put, Post, &19 &\url{https://petstore3.swagger.io/api/v3}\\
& Get, Delete & & \\\hline
Bills &Get &19 &\url{https://bills-api.parliament.uk}\\\hline
Canada Holidays &Get &4 &\url{https://canada-holidays.ca}\\\hline
Cat Fact &Get &3 &\url{https://catfact.ninja}\\\hline
\end{tabular}
\end{scriptsize}
\end{table}

Two LLMs were used in this experiment: GPT-4o and DeepSeek (DeepSeek-Reasoner $V3.1$). To ensure reproducibility and reduce variance in the generated results, all LLM requests were made with fixed parameters. Specifically, the temperature on GPT-4o was set to $0.6$, and DeepSeek was set to $0.0$ without a $max\_token$ limitation. All experiments were conducted on a machine with an Apple M2 chip, running macOS 14.5 (23F79) with 16 GB of memory.

\subsection{Experiment Process}

The experiment was conducted by one human tester with 5 years of industry experience of testing RESTful APIs to ensure the consistency of the evaluation experiment and the meaningfulness of human judgement and assessment of the quality of LLM generated entities. It proceeded through the following steps for each API of the benchmark one by one using the system in the workflow described in Section \ref{sec:WorkflowAndArchitecture}. 

\begin{itemize}
\item \emph{API Parsing}. The API operation details were parsed from the Swagger specification in JSON format loaded directly from their URLs, which includes operation summary, description, paths, methods, parameters, and responses. Schema place holders were replaced with actual data contained in the API Swagger file.
\item \emph{Scenario Generation and Review}. Test scenarios were generated by the selected LLM from the parsed API operation details. The tester reviewed each generated scenario to determine whether it should be accepted, modified, or rejected, and whether new scenarios should be added to the list. The decisions are made according to the following criteria. 
\begin{itemize}
\item \emph{Accept}. The generated scenario accurately captured the operation's input parameters and responses.
\item \emph{	Modify}. The generated scenario was mostly correct but required adjustments to ensure accuracy.
\item \emph{Reject}. The generated scenario was invalid, irrelevant, meaningless or duplicate. 
\item \emph{Add}. A new scenario was added by the tester when the scenario was missing.
\end{itemize}
The actions taken by the human tester in this step were recorded and later used to calculate scenario coverage and operation coverage as defined in subsection \ref{sec:Metrics}.
\item \emph{Script Generation and Review}. The executable test scripts were generated by the selected LLM from the parsed API operation details and scenarios after the scenarios were reviewed and accepted. Testers then reviewed the test scripts to verify whether they correctly implemented the scenario logic and/or API interactions, and whether the script needed to be modified. 

A test script is modified if it is mostly correct but required adjustments to ensure accuracy or executability. Syntax correctness, data type correctness, and method coverage (defined in subsection \ref{sec:Metrics}) were measured after test scripts were generated. These metrics were also updated after a modification of the test script. Usability was also quantified by the number of characters modified to make the script executable and logically correct.
\item \emph{Execution of Test Script and Analysis of Test Results}. Test scripts were executed through the Test Executor agent, which runs the Pytest command. The actual responses returned by the web service were captured and stored in the database and displayed together with the test script, if an error is detected. These execution results were later used to measure status code coverage according to the formula given in subsection \ref{sec:Metrics}.
\item \emph{Review and Validation of Quality Metrics}. For quality metrics evaluated by LLM, i.e. data type correctness, method coverage, and status code coverage, the human tester reviewed the results to confirm whether the LLMs' assessments were accurate.
\end{itemize}

\subsection{Performance Evaluation Metrics}\label{sec:Metrics}

A number of metrics are used to evaluate the performance of LLMs on their capability of performing various tasks in the process of API testing.  These metrics measure the ability of LLMs to produce accurate, diverse, and usable outputs across both scenarios and scripts. In addition, quality metrics are introduced to assess how effectively LLMs can verify data types, API methods, and response status codes. 
To ensure reproducibility and transparency, all metrics were calculated based on objective data collected implicitly in the testing process. For metrics requiring LLM assistance (such as method coverage, data type correctness, and status code coverage), the values generated by a LLM were reviewed by the tester, and then stored in the database. The metrics were then computed through SQL aggregation queries as described below. 

\emph{Syntax Correctness}. Each generated test script was parsed using the ast.parse() function in Python's Abstract Syntax Tree (AST)\footnote{AST — Abstract Syntax Trees: \url{https://docs.python.org/3/library/ast.html} (Accessed: 28 August 2025)} module. If parsing succeeded without raising a syntax error, the script was marked as correct. Otherwise, it was marked as incorrect. The syntax correctness is the proportion of test scripts that are syntactically correct. 

\begin{equation}
Cor_{Syn}(api) =\frac{|\{t \in T_{LLM}(api)|Valid_{Syn}(t)\}|}{|T_{LLM}(api)|}
\end{equation}
where $Valid_{Syn}(t)$ means test script $t$ is syntactically correct, $T_{LLM}(api)$ is the set of test scripts generated by LLM for the API $api$. $|X|$ is the number of elements in the set $X$ , e.g. $|T_{LLM}(api)|$  is the number of test scripts in $T_{LLM}(api)$. 

\emph{Data Type Correctness}. The data type correctness of a test script is evaluated by LLM with API operation details, scenarios, and generated test scripts as the input. The LLM's evaluation of each test script is reviewed and confirmed or corrected by the human tester. Data Type Correctness as a metric on the quality of LLM generated test scripts is the proportion of test scripts that are data type correct over all test scripts, which is formally defined as by the following formula.
\begin{equation}
Cor_{DT}(api) =\frac{|\{t \in T_{LLM}(api)| Valid_{DT}(t)\}|}{|T_{LLM}(api)}
\end{equation}
where $Valid_{DT}(t)$ means that $t$ is a valid Pytest script and correct in data types. 

\emph{Usability}. For each test script, the Levenshtein character level distance between the test script generated by a LLM and its final executed version was calculated using Levenshtein.distance() function in the Python's Levenshtein library\footnote{Levenshtein: \url{https://pypi.org/project/Levenshtein/} (Accessed: 22 September 2025)}. The usability of the test scripts for an API is the average editing distance over all test scripts of the API. 

\begin{equation}
Usability(api) =\frac{\sum_{t \in T_{LLM}(api)} \|t,t'\|_L }{|T_{Fin}(api)|} ,
\end{equation}
where $T_{Fin}$ is the set of final test scripts after review and correction of LLM generated test scripts. For each $t \in T_{LLM}(api)$, $t' \in T_{Fin}(api)$ is the final test script obtained by modifying $t$, $\|t,t'\|_L$ is the Levenshtein distance between them. The smaller Levenshtein distance means less editing is required on the generated test script before it can be used. Therefore, the less the value of usability metric means better usability. 

\emph{Unit Test Scenario Coverage}. It is defined as the proportion of scenarios LLM generated for an operation accepted without any modification to the total number of scenarios finally used to test an API operation. Let $op \in Ops(api)$ be a given operation of the API, and  $S_{Fin}(op)$ be the final set of scenarios for unit test of API operation $op$, $S_{LLM}(op)$ be the set of scenarios generated by LLM for API operation $op$. The LLM's scenario coverage for $op$, denoted by $Cov_{US}(op)$, is formally defined by the following formula.  

\begin{equation}
Cov_{US}(op) = \frac{|S_{LLM}(op) \cap S_{Fin}(op)|}{|S_{Fin}(op)|}
\end{equation}

The overall scenario coverage over all operations of the API is defined by the following formula. 
\begin{equation}
Cov_{US}(api)= \frac{|\cup_{op\in Ops(api)} \left(S_{LLM}(op) \cap S_{Fin}(op)\right)|}{|\cup_{op \in Ops(api)}S_{Fin}(op)|}
\end{equation}

\emph{System Test Scenario Coverage}. Each system test scenario generated by LLM was reviewed and validated manually by the human tester to confirm the correct ordering of operations and the data parameters passed between the operations. If there is any error, the human tester will make modifications or completely delete the scenario. If the set of system test scenarios are inadequate, new system test scenarios can also be added manually. The system test scenario coverage is the proportion of system test scenarios generated by LLM and accepted without any modification in the set of finally used system test scenarios for the API. The metric is formally defined as follows. 

\begin{equation}
Cov_{SS}(api)=\frac{|S_{Fin}^{Sys}(api) \cap S_{LLM}^{Sys}(api)|}{|S_{Fin}^{Sys}(api)|} 
\end{equation}
where $S_{Fin}^{Sys}(api)$ is the final set of scenarios used as system test scenarios for the API, $S_{LLM}^{Sys}(Api)$  is the system test scenarios generated by LLM and accepted by the tester without any change. 

\emph{Operation Coverage}. It is concerned with how well the operations in the API are covered by system test scenarios. It is defined as the proportion of API operations included in the system test scenarios and their corresponding test scripts. 

\begin{equation}
Cov_{Ops}(api)=\frac{|Ops(T_{Fin}(api))|}{|Ops(api)|} 
\end{equation}
where $Ops(T)$ is the set of API operations covered by test cases in the set $T$ of test scripts. Formally, 
\[Ops(T) = \{op \in Ops(api) ~|~ \exists t \in T.(t ~contains~ op)\} \]
 
\emph{Status Code Coverage}. It is concerned with the status codes in the service response messages returned by API operation requests. The status codes are collected from the responses received after executions of the test scripts for each operation. Status code coverage measures how well the set of possible status codes are covered by test executions in terms of the proportion of status codes contained in the received service response messages. It is a test adequacy measure widely used in API testing practice.

\begin{equation} 
Cov_{SCode}(op)=\frac{|StutusCode_{Rec} (op)|}{|StatusCode_{Exp} (op)|}
\end{equation}
where $StutusCode_{Rec}(op)$ is the set of status codes received after executing on all test scripts for testing operation op, $StatusCode_{Exp}(op)$ is the set of status codes that could be returned by the API operation $op$. 

\subsection{Experiment Results}

\subsubsection{Unit Test Scenario Coverage}

Table \ref{tab_4} presents the percentage of valid scenarios generated by GPT-4o and DeepSeek V3.1 across the five APIs. 

\begin{table}[htb]
\caption{Unit Test Scenario Coverage}
\label{tab_4}
\centering
\begin{tabular}{|l|c|c|}
\hline
\textbf{API} &\textbf{GPT-4o}&\textbf{DeepSeek}\\\hline
Car &97\%&98\% \\\hline
Petstore3&100\%&100\% \\\hline
Bills&95\%&91\% \\\hline
Canada Holidays&78\%&100\% \\\hline
Cat Fact&100\%&100\% \\\hline
\textbf{Average} &\textbf{94\%} &\textbf{98\%} \\\hline
\end{tabular}
\end{table}

Both models achieved more than 90\% scenario coverage for almost all APIs. The only exception was GPT-4o on the Canada Holidays API, where the coverage dropped to 78\%. The main reason was that some request parameters with Enum or list values, as well as certain boundary conditions, were not covered. On the Bills API, DeepSeek scored 91\%, which is the lowest scenario coverage for DeepSeek. 

Overall, both models achieved high average unit test scenario coverage over the five APIs. This demonstrates that both LLMs are capable of performing the unit test scenario generation task. Comparing the LLMs, DeepSeek was slightly better with a score of 98\% than GPT-4o (94\%) by generating more complete scenarios from the API specifications. 

\subsubsection{System Test Scenario Coverage}

System Test Scenario Coverage metric measures LLM's capability of generating system test scenarios in the form of operation call sequences. This is one of the most challenging tasks in testing APIs as the research on automation of REST API tests has demonstrated; see Section \ref{sec:Introduction}. 

Table \ref{tab_5} presents the results on system test scenario coverage. For the Bills API, GPT-4o and DeepSeek scored only 50\% and 40\%, respectively. This was mainly due to missing operation sequences, for example, missing follow-up operations after obtaining a bill Id. In some cases, the generated scenarios failed to include at least two operation calls as required by the prompt, producing only a unit test scenario for individual operations. Similar issues occurred with the Car API, where some valid sequences of operations were missed, and several invalid sequences were generated. In the Petstore API, the low accuracy was also due to invalid sequences.

\begin{table}[htb]
\caption{System Test Scenarios Coverage}
\label{tab_5}
\centering
\begin{tabular}{|l|c|c|}
\hline 
\textbf{API} &\textbf{GPT-4o}&\textbf{DeepSeek}\\\hline
Car &70\%&75\% \\\hline
Petstore&76.47\%&75\% \\\hline
Bills&50\%&40\% \\\hline
Canada Holidays&100\%&100\% \\\hline
Cat Fact&100\%&100\% \\\hline
\textbf{Average}&\textbf{79\%} &\textbf{78\%} \\\hline
\end{tabular}
\end{table}

Overall, both LLM models performed better on APIs with fewer operations. The performance declined with the increase in the number of operations in the APIs. 

Comparing the performances of LLMs, the difference between GPT-4o and DeepSeek was minimal. 

Given the fact that generating meaningful sequences of API operation calls is one of the most challenging tasks, LLMs performed far better than we originally expected. 

\subsubsection{Bug Detection Ability}

\begin{figure}[htb]
\centering
\includegraphics[width=3.4in]{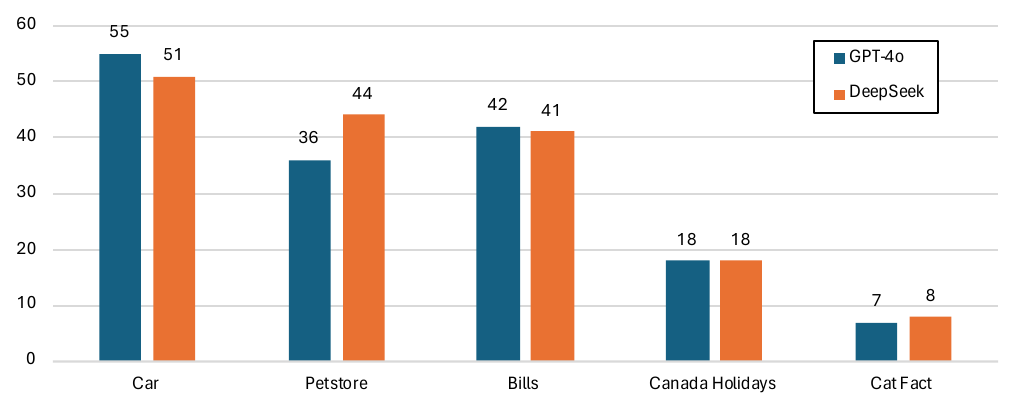}
\caption{Number of Bugs Detected by GPT-4o and DeepSeek.}
\label{fig_5}
\end{figure}

The test scripts generated by LLMs contain assertions to check the correctness of the responses from the web services. This enables bugs of the web services to be detected if a response fails the assertion. Table \ref{tab_6} shows the number of bugs discovered across five APIs. The data is also visualised in Fig. \ref{fig_5}. 

These bugs include service functional errors, inconsistencies between actual service responses and the API specification, and undefined response status codes. Both LLMs detected a substantial number of such issues across all APIs and performed similarly in bug detection. 
 
\begin{table}[htb]
\caption{Numbers of Bugs Detected}
\label{tab_6}
\centering
\begin{tabular}{|l|c|c|}
\hline 
\textbf{API} &\textbf{GPT-4o}&\textbf{DeepSeek}\\\hline
Car &55&51 \\\hline
Petstore3&36&44 \\\hline
Bills &42&41 \\\hline
Canada Holidays &18&18 \\\hline
Cat Fact &7&8 \\\hline
\textbf{Total} &\textbf{158}&\textbf{162} \\\hline
\end{tabular}
\end{table}

It is worth noting that, as shown in Table \ref{tab_7}, MASTEST's performance remains at the same level in terms of the number of bugs detected per API operation while the scale of the APIs increases, where the scale is measured by the number of API operations. 

\begin{table}[htb]
\caption{Number of Bugs Detected Per Operation}
\label{tab_7}
\centering
\begin{tabular}{|l|c|c|c|}
\hline 
\textbf{API} & \textbf{\#Ops} &\textbf{GPT-4o}&\textbf{DeepSeek}\\\hline
Car&24&2.29&2.13 \\\hline
Petstore&19&1.89&2.32 \\\hline
Bills &19&2.21&2.16 \\\hline
Canada Holidays&4&4.50&4.50 \\\hline
Cat Fact&3&2.33&2.67 \\\hline
\end{tabular}
\end{table}

Moreover, in terms of the number of bugs detected per operation, MASTEST performed equally well even when the scenario coverage is less satisfactory. For example, both GPT-4o and DeepSeek did well on detecting bugs of the Bills API while they only achieved 50\% and 40\% system test scenario coverage, respectively. 

\subsubsection{Quality of Generated Test Scripts}

In terms of the quality of the test scripts generated by LLMs, both GPT-4o and DeepSeek achieved 100\% on syntactical correctness as all test scripts were syntactically correct. 

For data type correctness, the experiment data is presented in Table \ref{tab_8}. The data shows that DeepSeek outperformed GPT-4o on all five APIs, with particularly large gaps on Bills (81.50\% vs. 64.44\%) and Petstore (86.74\% vs. 68.67\%). Over all five APIs, DeepSeek achieved the average data type correctness of 87.84\%, while GPT-4o is only 74.56\%. 
 
\begin{table}[htb]
\caption{Data Type Correctness}
\label{tab_8}
\centering
\begin{tabular}{|l|c|c|}
\hline 
\textbf{API} &\textbf{GPT-4o}&\textbf{DeepSeek}\\\hline
Car&86.51\%&92.22\% \\\hline
Petstore&68.67\%&86.74\% \\\hline
Bills&64.44\%&81.50\% \\\hline
Canada Holidays&72.41\%&92.64\% \\\hline
Cat Fact&80.77\%&86.11\% \\\hline
\textbf{Average}&\textbf{74.56\%}&\textbf{87.84\%} \\\hline
\end{tabular}
\end{table}

\subsubsection{Usability}

Usability measured by the average editing distance from LLM Generated test scripts to the test scripts actually used over all test scripts. The results varied across APIs. On average, DeepSeek generated test scripts requiring less manual editing for the Car and Canada Holiday APIs, whereas GPT-4o required fewer edits for Petstore, Bills, and Cat Fact APIs. Table \ref{tab_9} summarises the average editing distance required for the generated scripts. The data is also visualised in Figure 6. 

\begin{table}[htb]
\caption{Usability Measured by Average Editing Distance}
\label{tab_9}
\centering
\begin{tabular}{|l|c|c|}
\hline 
\textbf{API} &\textbf{GPT-4o}&\textbf{DeepSeek}\\\hline
Car&44.27&15.4 \\\hline
Petstore&14.85&40.27 \\\hline
Bills&9.29&31.97 \\\hline
Canada Holidays&63.2&10.45 \\\hline
Cat Fact&20.12&25.02 \\\hline
\textbf{Average} & \textbf{30.35} &\textbf{24.62} \\\hline
\end{tabular}
\end{table}
 
\begin{figure}[htb]
\centering
\includegraphics[width=3.4in]{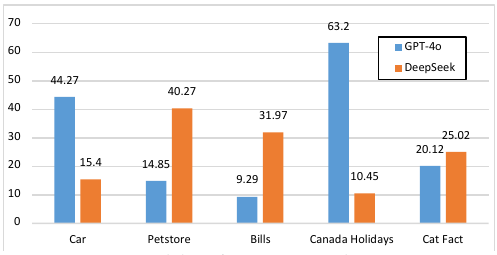}
\caption{Usability of LLM Generated Test Scripts.}
\label{fig_6}
\end{figure} 

\subsubsection{Test Adequacy}

The test adequacy is measured by two metrics: Operation coverage for the proportion of operations in the API are actually tested by the test scripts corresponding to system test scenarios, and Status Code Coverage for the proportion of status codes exercised by testing. 

Table \ref{tab_10} presents the data on operation coverage for different APIs. GPT-4o performed extremely well by achieving 100\% on all APIs, while DeepSeek V3.1 was only slightly lower on some APIs (92.22\% on Car, 98.89\% on Petstore). 

\begin{table}[htb]
\caption{Operation Coverage of System Tests}
\label{tab_10}
\centering
\begin{tabular}{|l|c|c|}
\hline 
\textbf{API} &\textbf{GPT-4o}&\textbf{DeepSeek}\\\hline
Car &100\%&92.22\% \\\hline
Petstore3&100\%&98.89\% \\\hline
Bills&100\%&99\% \\\hline
Canada Holidays&100\%&100\% \\\hline
Cat Fact &100\%&100\% \\\hline
Average&100\%&98\% \\\hline
\end{tabular}
\end{table}

Table \ref{tab_11} gives the data on Status Code Coverage.  The data shows that DeepSeek again has a higher score than GPT-4o, achieving 95.50\% coverage of status codes on Bills and 87.29\% on Petstore, compared to GPT-4o's 27.78\% and 52.67\%. Over all five APIs of the benchmark, DeepSeek achieved an average of 83.83\%, while GPT-4o is only 54.64\%. 

\begin{table}[htb]
\caption{Status Code Coverage}
\label{tab_11}
\centering
\begin{tabular}{|l|c|c|}
\hline 
\textbf{API} &\textbf{GPT-4o}&\textbf{DeepSeek}\\\hline
Car &43.82\% &85.86\% \\\hline
Petstore&52.67\% &87.29\% \\\hline
Bills&27.78\% &95.50\% \\\hline
Canada Holidays&75.86\% &83.82\% \\\hline
Cat Fact &73.08\% &66.67\% \\\hline
Average&54.64\%&83.83\% \\\hline
\end{tabular}
\end{table}

In summary, both GPT-4o and DeepSeek achieved remarkable performance on all quality metrics. This indicates that the proposed multi-agent system empowered by LLMs is practically usable. 

\begin{figure}[htb]
\centering
\includegraphics[width=3.4in]{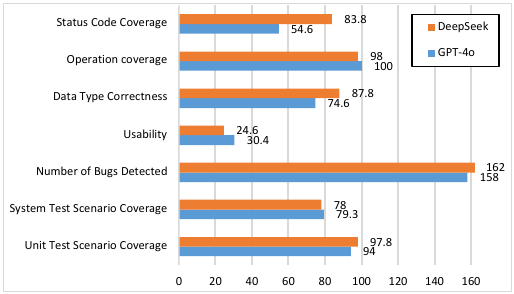}
\caption{Performance Overview.}
\label{fig_7}
\end{figure}

Comparing GPT-4o with DeepSeek, as shown in Fig. \ref{fig_7}, GPT-4o performed better on operation coverage, usability and system test scenario coverage, while DeepSeek outperformed in status code coverage, data type correctness, bug detection, and unit test scenario coverage. However, the differences in their performances are very small. 
 
\section{Related Work}\label{sec:RelatedWorks}

This section reviews related works and compares them with the method proposed in this paper. We will focus on the applications of LLMs in the testing of RESTful APIs. 

According to the source of information that LLMs are applied to, the work on application of LLMs to REST API test can be classified into two categories. 
\begin{enumerate}
\item \emph{White-Box approach}, which takes program code under test as input and utilises LLMs' capability of understanding of program code to perform testing activities; 
\item \emph{Black-Box approach}, which does not apply LLM to program code. 
\end{enumerate}

Almost all the work reported in the literature belongs to the black-box approach. One of the rare exceptions is the work by Rincon et al. in 2025 \cite{rincon2025llm}, which employs LLMs to generate JUnit test sets for integration testing of REST APIs with both program source code and OpenAPI specifications as inputs. 

Works in the black-box approach, to which our work also belongs, vary on two aspects: (a) how LLMs are utilised, (b) the structure of workflow and the architecture of multi-agent systems. 

\subsection{Utilisation of LLMs}

According to the specific capability of LLMs utilised, existing works can be classified into three categories: (a) those utilising the capabilities of NLP, (b) those utilising the capability of code generation, and (c) those utilising the capabilities of directly performing software engineering tasks such as test design, etc. In the following, we review existing works in each category, respectively.

\bulletList{Utilisation of NLP Capability} 

As Kim et al. pointed out \cite{kim2023enhancing}, one of the main weakness in conventional algorithmic techniques and methods for testing RESTful APIs is that they are unable to process the natural language descriptions of API operations and parameters as well as the meanings of the names of operations and parameters in the context of the real world. Although conventional NLP techniques have been explored, e.g. in the works of NLPtoREST by Kim et al. \cite{kim2023enhancing}, LLMs offer a much powerful capability to process such data, and therefore, can be used to enhance existing algorithmic methods, such as fuzzing. This is explored by a number of researchers. The most well-known works in this category include RESTGPT \cite{kim2024leveraging}, RESTLess \cite{zheng2024restless}, and KAT \cite{le2024kat}, the work by Liu et al. \cite{liu2026rest}, etc. These works differ from ours that we feed the whole API specification to LLMs to understand the semantics of the specification in the context of the real world and then to generate other test artefacts directly. In contrast, they extract the natural language descriptions out of the specification. 

In contrast to the above methods that reply on the existence of an REST API specification, RESTSpecIT proposed by Decrop et al. \cite{decrop2024you} infers the OpenAPI specification and conducts testing without any documentation of APIs. Again, different from our system, RESTSpecIT works like a fuzzer, but employs GPT 3.5 to generate mutant tests. 

\bulletList{Utilisation of Code Generation Capability}

In this approach, LLMs are employed to utilise its capability of code generation to generate test code or test scripts in addition to utilise LLMs' capability of NLP. 

A typical example of this type is APITestGenie proposed by Pereira et al. in 2024 \cite{pereira2024apitestgenie}. APITestGenie takes business requirements description in natural language together with REST API specifications as input to directly generate executable test scripts by invoking a LLM. To deal with the scalability problem in order to process large files of requirements documents and long REST API specifications, it applies the Retrieval Augmented Generation (RAG) techniques to reduce the number of tokens in the prompts. The generated test scripts are also improved through invocations of a LLM with the responses obtained from executing the test scripts. This approach relies heavily on high-quality business documentation and API specifications. They concluded that human intervention is needed to validate LLM's output. 

More recently, in 2025 Barradas et al. proposed RestTSLLM \cite{barradas2025combining}, a method that employs LLMs to automatically generate RESTful API test cases presented in the Test Specification Languages (TSLs) and then by employing a LLM with few-shot prompts again to generate executable test code. This method also requires high-quality TSL input. Moreover, although the reasons for why scripts failed are analysed, it does not verify whether successfully executed scripts are logically correct.

The above methods not only differ from our approach in the way LLMs are used in term of what are generated by LLMs, but also in how to address the oracle problem. Similar to traditional algorithmic fuzzers for REST API test, the above methods rely on status code in the response message to detect server errors and mismatching between expected and response status codes. Yet, they do not comprehensively cover all status codes. In contrast, the test scripts generated by MASTEST include assertions on the contents of the responses in addition to the assertions on status code. 

Huynh et al. \cite{huynh2024using} explored the possibility of utilisation of LLMs for generating test oracles. Their system APITesting generates test validation code to check the logical correctness of response data by utilising the capability of LLMs to interpret natural language descriptions of returned objects found in API specifications together with their schemas to mine constraints on response bodies and operations.  

\bulletList{Utilisation of Capabilities of Performing Test Activities} 

In this approach, a wider range of the LLMs' capabilities is utilised to performing various software testing activities directly, such as design test cases. 

An early attempt of this approach is the PULSE-UI system developed by Leu et al in 2024 \cite{leu2024reducing}. PULSE-UI uses a LLM in generating, analysing and manipulating test scenarios for REST API. It also provides a web-based user interface for visualisation of test scenarios, API endpoints, creating and linking test sequence and automating test data generation. A preliminary user study of the tool demonstrated its ability to reduce workload on human testers, enhance test quality and improve user experience in testing REST APIs. However, no qualitative experiment data were reported. 

In 2025, Zhang et al. proposed LogiAgent \cite{zhang2025logiagent}, an LLM-driven multi-agent system for automating logic testing of RESTful systems. Three LLM-based agents were were developed, which are Test Scenario Generator, API Request Executor, and API Response Validator. A Scenario Scheduler controls the executions of these agents in an iterative process. Each cycle starts with an invocation of the Test Scenario Generator agent to generate a new test scenario in the form of a sequence of API requests paired with a test oracle defining the expected response. It then invokes the API Request Executor to construct request payloads (values for API parameters) of the test case and then calls a function to execute the request and collect the response from the API. Finally, the API Response Validator is invoked to assess the response from the API according to the expected behaviour in the scenario. To ensure each cycle of the loop generates a new scenario, the scenarios executed in the previous cycles are stored in the Execution Memory together with the submitted request payloads and the responses received from API in each step of the scenarios. The information of past executed scenarios are retrieved and used in the next cycle of scenario generation. The method is similar to our approach in the way LLMs are used directly to generate test scenarios. However, it differs from our approach that we ask LLM to generate a comprehensive set of diverse scenarios while LogiAgent generates scenarios one by one depending on a complicated external mechanism, which includes the Execution Memory and the construction of a so called the API Relationship Graph (ARG), an undirected graph constructed from API documentation that encodes semantic relationships between APIs. Second, we generate test scripts that are human readable for validation and modification and are reusable for regression testing of the REST API. Experiment results show that LogiAgent can effectively detect 234 logical issues, achieving an accuracy of 66.19\%. Despite these strengths, this approach to generating new scenarios depends on the knowledge of the previous scenarios, limiting test parallelism and scalability. 

Our method has similarity to LogiAgent in the way that LLMs are utilised to directly perform generative testing activities in test scenario designs and test script generations. However, our generated test scripts not only contains API invocations with parameter values, but also oracle code that asserts the correctness of the responses received from the web services. 

\subsection{Workflow And Architecture of Multi-Agent Systems}

MASTEST is a multi-agent system that employs LLM-based intelligent agents as well as programmed agents and human agents to implement various testing activities that form a complete workflow that covers the whole process of RESTful API tests. 

The notion of multi-agent systems have been employed in the research on REST API testing tools explicitly such as in AutoRestTest \cite{kim2025multi, stennett2025autoresttest} and LogiAgent \cite{zhang2025logiagent}, and implicitly with multiple invocations of LLMs with different prompts and for different purposes in one system, such as in PULSE-UI \cite{leu2024reducing}, APITestGenie \cite{pereira2024apitestgenie}, RESTSpecIT \cite{decrop2024you}, etc. Our MASTEST system differs from them not only in the roles assigned to LLM-based intelligent agents as discussed above, but also in the structure of the workflow. The MASTEST's workflow is the complete workflow of current based practice in the industry. In contrast, AutoRestTest \cite{kim2025multi, stennett2025autoresttest} and LlamaRestTest \cite{kim2025llamaresttest} used a fuzzer-like framework, LogiAgent \cite{zhang2025logiagent} is an iterative process, while the workflows of PULSE-UI \cite{leu2024reducing} and APITestGenie \cite{pereira2024apitestgenie} is not explicitly stated, over simplified, especially, omitted quality assurance tasks. 

MASTEST's architecture and the corresponding workflow model are not a result of trial-and-error process of exploration. As presented in Section \ref{sec:WorkflowAndArchitecture}, it is the result of a careful design of the workflow structure following the design principles for role-based multi-agent systems and assigning roles to various kinds of agents following the design rules that address the common problems in the development of LLM applications with the objective of reducing workload on human testers while maintaining high work quality. This distinguishes MASTEST from all existing works in the applications of LLMs to REST API test, where few reported how the system was designed. 

\section{Conclusion}\label{sec:Conclusion}

This section concludes the paper with a summary of our main contributions and a discussion of the limitations of the MASTEST system and the directions of future work.

\subsection{Main Contributions} 

The main contributions of the paper are three-fold. 

First, we proposed a new LLM based agentic approach to automating REST API test and developed a multi-agent system called MASTEST. In this approach, intelligent agents empowered by LLMs directly perform generative testing activities such as test scenario designs, test script generation. Programmed agents perform algorithmic routine tasks such as parsing Swagger specifications of REST APIs, checking the syntax correctness of generated test scripts, and invoking Web services, and so on. Human agents participate in the testing process through GUIs to perform quality assurance tasks such as reviewing the output of intelligent agents and correcting their errors. 

Second, we proposed a set of design principles for architectural design of multi-agent systems that integrates LLM-based intelligent agents with programmed agents and human agents in a workflow the reflects the current best practice in the industry. Its application is demonstrated by the design of MASTEST system. 

Finally, we conducted an experiment to demonstrate the effectiveness of the proposed MASTEST system and its practical usability. A set of performance metrics tailored to evaluate multi-agent systems were formally defined and applied in the experiment to measure not only the overall performance of the whole system but also the performances of each individual LLM-based intelligent agent. 

\subsection{Limitations and Future Works}

The MASTEST system and the experiments are still preliminary. There are several limitations, which could be further improved. 

First, due to resource limitation, we have conducted experiments with only five APIs and two LLMs. Results may vary with other LLMs or different versions of the same models. It is worth conducting further experiments with more LLMs and more APIs. Moreover, it is possible to use different LLMs to realise different intelligent agents. Another issue that is worth consideration is the design of the prompts. Currently, for each intelligent agent, one prompt template is used. It is worth exploring if different LLMs should use different prompt templates to achieve the best performance. 

Second, the experiment focused on the quality of LLM generated test scenarios and scripts. The former is measured by unit test scenario coverage and system test scenario coverage, API operation coverage and status code coverage. The latter is measured by data type correctness, syntax correctness, and semantics correctness. Other aspects, such as back-end code coverage, were not considered. It is worth further investigation. 

Moreover, while MASTEST covers the whole process of API test, its functionality could be further improved. For example, when generating system test scenarios and system test scripts for the entire API with multiple operations, the input to LLM may exceed the model's limit on the number of tokens. Our current solution is to allow manually select a subset of operations. We are exploring alternative approaches to generate system test scenarios to reduce human efforts. Another functionality that could be improved is the review tasks, which are currently manually conducted by human testers. Although the current system can significantly reduce the overall manual workload, a future work is to explore how to further reduce the workload on human users, for example, by employing LLMs to assist reviewing activities. It would be even more interesting if the review results could be feedback to the agents to improve their performances. 

Finally, test data generated by LLMs could be invalid or unrealistic in real business contexts. A future work is to integrate database and real-world data as inputs, even to employ LLMs to generate realistic test data. 

\bibliographystyle{IEEEtran}  
\bibliography{MASTESTreferences}
%\end{thebibliography}

\end{document}